\documentclass[aps,prx,reprint,amsmath,amssymb,floatfix,longbibliography]{revtex4-2}

\usepackage{graphicx}
\usepackage{dcolumn}
\usepackage{bm}
\usepackage[english]{babel}
\usepackage{soul,multirow,booktabs,comment}
\usepackage{enumitem}
\usepackage{array}
\usepackage{tabularx}
\newcommand{\my}[1]{\textcolor{black}{#1}}

\newcommand{\myg}[1]{\textcolor{black}{#1}}

\usepackage{hyperref}
\hypersetup{
	colorlinks = true,
	linkcolor = blue,
	citecolor = blue,
	urlcolor = blue,
	filecolor = blue
}

\begin{document}


\title{Fluorescence quenching in plasmonic dimers due to electron tunneling}

\author{Henrikh M. Baghramyan$^{1}$}
\author{Cristian Cirac\`{i}$^{1}$}
 \email{cristian.ciraci@iit.it}

\affiliation{$^{1}$Center for Biomolecular Nanotechnologies, Istituto Italiano di Tecnologia, Via Barsanti 14, 73010 Arnesano (LE), Italy}

\date{\today}

\begin{abstract}
\noindent Plasmonic nanoparticles provide an ideal environment for the enhancement of fluorescent emission. On the one hand, they locally amplify the electromagnetic fields, increasing the emitter excitation rate, and on the other hand, they provide a high local density of states that accelerates  spontaneous emission.
However, when the emitter is placed in close proximity to a single metal nanoparticle, the number of nonradiative states increases dramatically, causing the fluorescence to quench.
It has been predicted theoretically that, through a judicious placing of the emitter, fluorescence in plasmonic nanocavities can be increased at monotonically. In this article, we show that such monotonic increase is due to the use of local response approximation in the description of the plasmonic response of metal nanoparticles.
We demonstrate that  taking into account the electron tunneling and the nonlocality of the surrounding system via the quantum hydrodynamic theory results eventually in a quenching of fluorescence enhancement also when the emitter is placed in a nanocavity, as opposed to  local response and  Thomas-Fermi hydrodynamic theory results. 
This outcome marks the importance of considering the quantum effects, in particular, the electron tunneling to correctly describe the emission effects in plasmonic systems at nanoscale. 
\end{abstract}


\maketitle
\section{Introduction}\label{sec01}
\noindent A quantum emitter in an excited states can undergo a decay by emitting a photon into one of the modes provided by the surrounding electromagnetic environment \cite{purcell,goy1983}.
Due to plasmonic effects, metal nanoparticles (NPs) can provide a high local density of electromagnetic states in their vicinity.
The emission of a quantum emitter than can be highly amplified at the proximity of plasmonic NPs \cite{tam2007}.
Similarly, the fields of an external electromagnetic wave can get localized and enhanced at the  metal NP surface well beyond the diffraction limit.
A quantum emitter placed in the vicinity of a metal NP then can also be more easily excited \cite{novotny_hecht_2012}.
These two processes -- the excitation and the emission -- contribute to the fluorescence of the emitter \cite{lakowicz2006}.
Plasmonic-enhanced fluorescence has several applications \cite{gaponenko2015}, spanning through biomedicine \cite{lakowicz2006a, lakowicz2006b, puerto2021}, sensors \cite{liSensors2015, langer2015}, quantum information \cite{FernandezDominguez:2018cw} and bionanotechnolgies \cite{darvill2013}. In particular, a hybrid plasmonic nanoantenna has been suggested in \cite{yang2020} for the efficient enhancement of emission directionality and keeping high quantum efficiency.
Also, more than 100 times fluorescence enhancements have been observed in \cite{cruz2020}, which is critical for readout by inexpensive point-of-care detectors.

At very short distances from a single metal NP, however, fluorescence quenches since the emitter couples mostly to nonradiative modes \cite{dulkeith2002,kuhn2006,novotny2006,bharadwaj2007}.
\my{To suppress plasmonic quenching, one could place the emitter in a nanocavity \cite{faggiani2015,kongsuwan2018,kongsuwan2018a,yang2020NatPhot,Cuartero-Gonzalez.2020,CuarteroGonzalez:ey}.}
\my{In the cavity formed by an NP coupled to a metal film, for example, quenching disappears because higher-order modes gain a radiative nature while the excitation rate increases \cite{kongsuwan2018a,yang2020NatPhot}.}
Within the \my{local response approximation (LRA),} it is predicted that, as the NP-film distance is reduced, the fluorescence rate monotonically increases, and quenching never occurs \my{\cite{kongsuwan2018a,yang2020NatPhot}.}
This, however, could be the result of a simplistic description rather than a physical consequence.
\my{In fact, local theories do not account for the nonlocal nature of the dielectric function of the metal at short distances \cite{ciraci_science,raza,goncalves2020}.}
Indeed, the hydrodynamic theory (HT) for the plasmonic response predicts bound to the maximum field enhancement  achievable in nanocavities \cite{ciraci_science}.
This is obtained by considering the electron pressure term when evaluating the permittivity of the metal. The pressure term originates from the Thomas-Fermi (TF) kinetic energy (KE) \cite{fermi, thomas}. 
However, the TF-HT is usually used with hard-wall boundary conditions at the metal surface, which does not allow electrons to spill outside of the metal \cite{raza11,ciraci2013a}. In addition, TF KE is only a local functional of the density, i.e. it depends on the value of the electron density at the point considered.

\myg{In Ref. \cite{Tserkezis:2017dx}, the authors consider the generalized nonlocal optical response (GNOR) by accounting for both, nonlocal response and surface-enhanced Landau damping,  and, in fact, predict a reduction of the fluorescence enhancement as compared to LRA calculations. However, even using the GNOR description there is no evidence of quenching.}
A more accurate approach would require taking into account the KE dependence on higher-order derivatives of the density.
The quantum hydrodynamic theory (QHT) adds such correction to the TF contribution, namely the von Weiz\"acker (vW) KE term \cite{vW}, which is equivalent to the quantum potential \cite{bohm1952, bohm1952a} and depends on the gradient of the electron density.
In the framework of TF and vW KE functionals, the spatial dependence of the electron density in the metal is naturally taken into account and spill-out, and electron tunneling effects can be easily considered in the calculation of plasmon energy \cite{toscano2015, ciraciQHT} while still retaining retardation effects \cite{ciraciQHT}. 
Nonetheless, the QHT in the TF-vW approximation may result in unstable resonances due to the propagation nature of the asymptotic solutions at high energies \cite{ciraciQHT, baghramyan2021}.
This behaviour can be amended by considering a Laplacian level correction in the KE \cite{fabio01, baghramyan2021}.

In this article, we apply the QHT to evaluate the fluorescence enhancement of the emitter in the gap of the Na jellium sphere dimer in the weak coupling regime.
\my{We show that for small enough gaps ( $\approx 0.5$~nm) the QHT predicts fluorescence quenching, as opposed to the LRA \cite{kongsuwan2018a,yang2020NatPhot}} calculations where the fluorescence enhancement monotonically increases as the gap closes.
Our formalism allows distinguishing the two quantum corrections that are often discussed in the literature: namely, spatial dispersion (or nonlocality) and electronic spill-out. In this paper, we show that the former does not produce any significant quenching \cite{Christensen:2014ha,Tserkezis:2016kp,Tserkezis:2017dx,jurga2017}, while the latter does. Therefore, both mechanisms manifest in fluorescence in a very different way.
Finally, we analyze the impact of Laplacian-level KE correction to the fluorescence spectra.

\section{Theory}\label{sec02}
\begin{figure}[!hbt]
\centering
\includegraphics[width=0.35\textwidth]{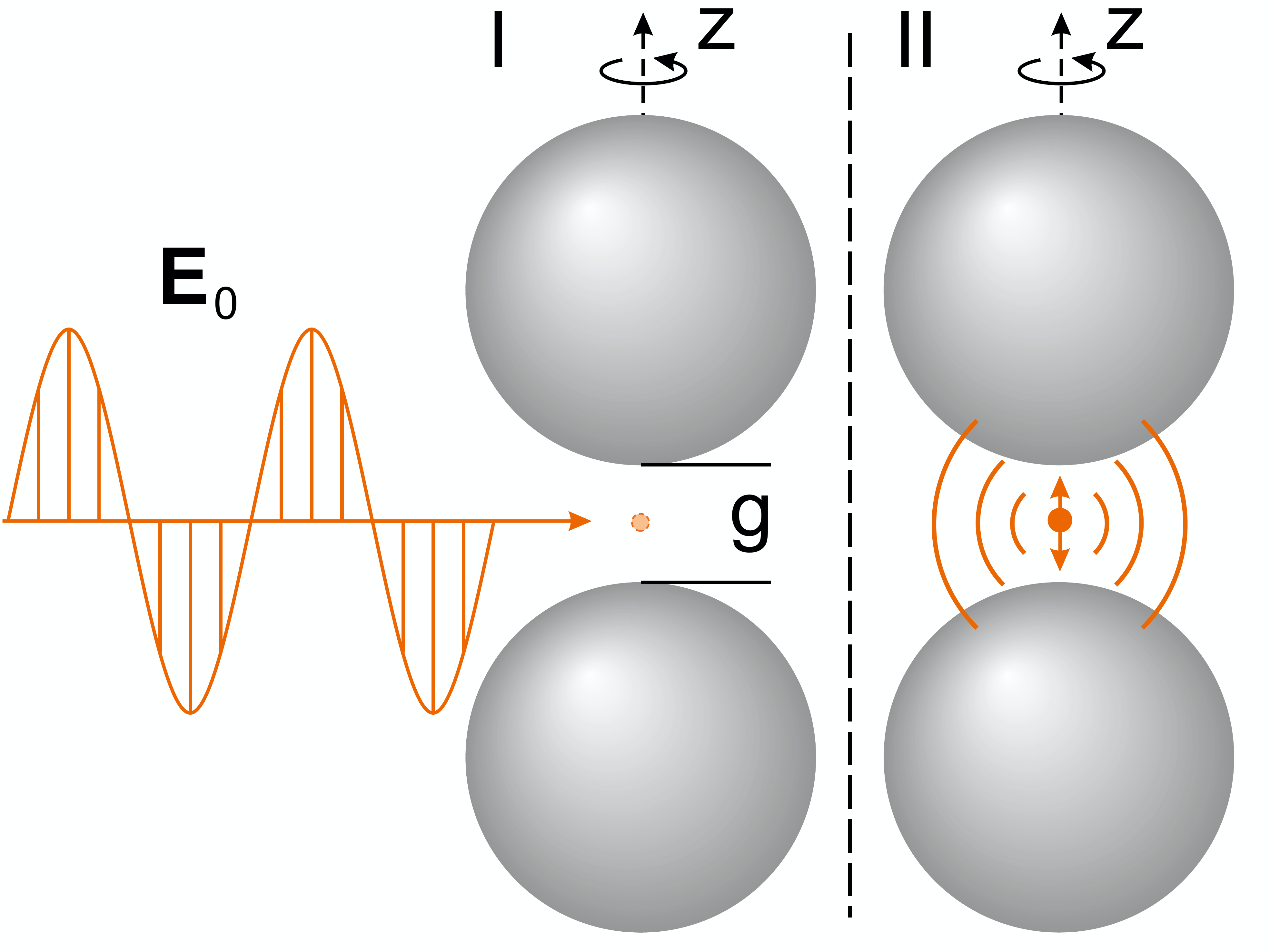}
\caption{Schematic image of the excitation and decay phases. 
First, the emitter is excited by a plane wave polarized along the $z$-axis impinging on the NP dimer (I); 
then, the emitter's dipole field interacts with the electromagnetic states provided by the dimer (II).}
\label{geom}
\end{figure}

\noindent To study the fluorescence process, the excitation of the emitter and its subsequent decay should be considered.
In the weak-coupling regime, the fluorescence enhancement, $\eta_{em}$, is the product of the excitation enhancement, $\eta_{ex}$, at the location of the emitter, $\mathbf{r}_{e}$, and quantum yield, $q$, of the emitter that accounts for the probability of the photons to couple to radiative states.
The excitation and the decay problems are solved independently, as schematically shown in Fig. \ref{geom}.

The first step then is to calculate the excitation enhancement of the incident radiation by the plasmonic environment. We consider an emitter placed at the center of a Na jellium sphere dimer at equal distances from the spheres.
Because the field in the dimer gap is mostly oriented along the $z$-direction, we only consider the case of an emitter vertically oriented.
Moreover, we assume a single frequency approximation and set the excitation and emission frequencies to be equal\cite{novotny2006}. At frequency $\omega$ and at the emitter location $\mathbf{r}_{e}$ $\eta_{ex}$ can be calculated by the following formula \cite{novotny_hecht_2012}:

\begin{equation}
\eta_{ex} = \frac{\gamma_{ex}}{\gamma_{ex}^0} =  \frac{\left|\left(\mathbf{E}_{\rm{I}}\left(\mathbf{r}_{e}, \omega\right)+\mathbf{E}_0\left(\mathbf{r}_{e}, \omega\right) \right) \cdot \hat{\mathbf{n}}_{e}\right|^{2}}{\left|\mathbf{E}_{0}\left(\mathbf{r}_{e}, \omega\right) \cdot \hat{\mathbf{n}}_{e}\right|^{2}}
\label{ext_enh}
\end{equation}

\noindent with $\gamma_{\mathrm{ex}}^{0}$ and $\gamma_{\mathrm{ex}}^{0}$ being the excitation rate in vacuum and in our system respectively, $\hat{\mathbf{n}}_{e}=\hat{\mathbf{z}}$ the dipole moment unit vector, $\mathbf{E}_{0}$ is the incident field (a plane wave polarized along the $z$-axis), and $\mathbf{E_{\rm{I}}}$ is the field scattered by the dimer.

The scattered field $\mathbf{E}_{\rm{I}}$ at the location $\mathbf{r}_{e}$ is obtained by solving the Maxwell equations coupled to the quantum hydrodynamic equation for the free electron dynamics in the Na NPs \cite{toscano2015,ciraciQHT}:

\begin{subequations}
    \begin{gather}
        \nabla\times\nabla\times\mathbf{E_{\rm{I}}} - \frac{\omega^2}{c^2}\mathbf{E_{\rm{I}}} = \omega^2\mu_0 \textrm{\textbf{P}}, \label{fq01a} \\
        \frac{en_0}{m_e}\nabla\left(\frac{\delta G\left[n\right]}{\delta n}\right)_1 + \left(\omega^2 + i\gamma\omega\right)\textrm{\textbf{P}} = -\varepsilon_0\omega_p^2\left(\mathbf{E_{\rm{I}}} + \mathbf{E}_{0}\right), \label{fq01b}
    \end{gather}
\label{fq01}
\end{subequations}

\noindent with $\varepsilon_0$ and $\mu_0$ are respectively the vacuum permittivity and permeability, $\textrm{\textbf{P}}$ is polarization vector, $c$ is the speed of light in free space, $e$ is the absolute value of the electron charge and $m_e$ is its mass, $\gamma$ is the damping rate, and $\omega_p\left(\textbf{r}\right) = \sqrt{e^2n_0\left(\textbf{r}\right)/\left(m_e\epsilon_0\right)}$ is the plasma frequency where $n_0\left(\textbf{r}\right)$ is the ground-state electron density. \myg{For simplicity, in this work, we neglect nonlocal  damping \cite{mortensen2014,ciraci2017visc,ciraciQEmmit,Tserkezis:2017dx}} and set the damping rate as $\gamma = 0.066~\textrm{eV} + v_\textrm{F}/R$ where $v_F$ is the Fermi velocity for Na and $R$ is the radius of the spheres.
The functional $G\left[n\right]$ accounts for the quantum nature of the electron gas total internal energy. It is the sum of noninteracting KE functional,  $T_s\left[n\right]$, and exchange-correlation (XC) functional, $E_{\textrm{XC}}^{\textrm{LDA}}\left[n\right]$, that is $G\left[n\right] = T_s\left[n\right] + E_{\textrm{XC}}^{\textrm{LDA}}\left[n\right]$.
$T_s\left[n\right]$ can be approximated considering the contributions of the TF, vW, Pauli-Gaussian second-order (PGS) and Laplacian-level functionals, as:

\my{\begin{equation}
    T_s\left[n\right] = \int\tau\left(n,w,q\right)\textit{d}^3\textrm{\textbf{r}},
\label{Ts_tau}
\end{equation}}

\noindent \my{where $w = \nabla n \cdot \nabla n$ and $q = \nabla^2 n$, and }

\begin{equation}
    \tau=\tau^{\mathrm{vW}}+\tau^{\mathrm{PGS}}+\tau^{\mathrm{TF}}\left[\beta q_{r}^{2}+2 \beta q_{0}^{2} \ln \left(1+q_{r} / q_{0}\right)\right]
    \label{tau}
\end{equation}

\noindent \my{In Exp. \eqref{tau}, each $\tau$ denotes the KE density associated with each functional \cite{baghramyan2021}.}

\begin{widetext}
    \begin{subequations}
        \begin{gather}
            \my{T_{\rm TF}[n] = \int\tau^{\rm TF}\left(n\right)\textit{d}^3\textrm{\textbf{r}}=\int\left(E_h a_0^2\right) \frac{3}{10} \left(3\pi^2\right)^{2/3}  n^{5/3}\textit{d}^3\textrm{\textbf{r}},}
            \label{tau_TF} \\
            \my{T_{\rm vW}[n, w] = \int\tau^{\rm vW}\left(n,w\right)\textit{d}^3\textrm{\textbf{r}}=\int\left(E_h a_0^2\right) 
            \frac{w}{ 8 n}\textit{d}^3\textrm{\textbf{r}},} \label{tau_vW} \\
            \my{T_{\rm PG\alpha}[n, w] = \int\tau^{\textrm{PG}\alpha}\left(n,w\right)\textit{d}^3\textrm{\textbf{r}} = \int\tau^{\rm TF}\left(n\right) e^{-\alpha C n^{-8/3}w}\textit{d}^3\textrm{\textbf{r}},} \label{tau_PG}
        \end{gather}
    \end{subequations}
\end{widetext}

\noindent \my{where $E_h$ is the Hartee energy, $a_0$ is the Bohr radius, $\alpha = 40/27$ for PGS, $\beta$ and $q_0$ are the free parameters of the Laplacian-level term and can be set as $\beta=0.25$ and $q_0=700$ as in \cite{baghramyan2021}, and  $q_r=3 \nabla^2 n/(40 \tau^{TF})$ is the reduced Laplacian. The expressions of the potentials in $\left(\delta G\left[n\right]/\delta n\right)_1 = \left(\delta E_\mathrm{XC}^{\mathrm{LDA}}/\delta n\right)_{1} + \left(\delta T_s/\delta n\right)_1$ can be found in the Section 2 in the SM. }

For the equilibrium density, we use the following model expression:

\begin{equation}
    n_0\left(\rho,z\right) =  \frac{f_0}{1 + e^{\kappa\left[r_{\uparrow}\left(\rho,z\right) - R\right]}}+\frac{f_0}{1 + e^{\kappa\left[r_{\downarrow}\left(\rho,z\right) - R\right]}},
    \label{dens}
\end{equation}

\noindent where $r_{\uparrow}\left(\rho,z\right) = \sqrt{\rho^2+\left(z-g/2 - R\right)^2}$ and $r_{\downarrow}\left(\rho,z\right) = \sqrt{\rho^2+\left(z+g/2 + R\right)^2}$ are for the upper and lower spheres respectively, and $g$ is the vertical separation measured from the surfaces of the spheres (see Fig. \ref{geom}). \my{Since we assume a jellium description, the surface of the ion background is located at a distance of $r_s$, the Wigner-Seitz radius, from the last atom such that a $g$ gap would correspond to a distance of $2r_s+g$ between the closest atoms.} In Exp. \eqref{dens} $\kappa = 1.05/a_0$ is fitted to the asymptotic decay of Kohn-Sham densities \cite{ciraciQHT}, with $a_0$  being the Bohr radius and $f_0$ is the normalization constant that is found from the condition:

\begin{equation}
    \int_{\Omega} n_{0} dV =2 \times \frac{4}{3} \pi R^{3} n^{+}
    \label{n0_norm}
\end{equation}

\noindent where $n^{+} = \left(4/3 \pi r_{s}^{3}\right)^{-1}/a_{0}^{3}$ denotes the density of positive charges with $r_s = 4$~a.u. Wigner-Seitz radius for Na.
The expressions for the KE functionals and linear potentials associated with them can be found in \cite{baghramyan2021, ciraciQHT}.

After calculating the plasmonic response of the dimer to the incident radiation, and hence the field that excites the emitter, we proceed to the second step that is the evaluation of the decay rates.

The rate of spontaneous emission is a sum of the radiative and nonradiative decays, i. e. $\gamma_{\rm sp}=\gamma_{\rm r}+\gamma_{\rm nr}$, and can be found from the relation:
\begin{equation}
  \gamma_{\mathrm{sp}}=\frac{2\pi^2 c}{\hbar \varepsilon_{0}}\rho_0 \left[{\mathbf{p}} \cdot \operatorname{Im}\{\mathbf{G}(\mathbf{r}_e, \mathbf{r}_e)\} \cdot {\mathbf{p}}\right],
  \label{gamma_sp}
\end{equation}
where $\hbar$ is the reduced Planck constant, $\rho_0 = \frac{\omega^2}{\pi^2 c^3}$ is the LDOS in vacuum, $\mathbf{p}$ is the transition dipole moment and $\bf G$ is the dyadic Green's function of the system  \cite{jurga2017}. 

The expression in Eq.~\eqref{gamma_sp} can be evaluated by using the relation:
\begin{equation}
    \mathbf{E_{\rm{II}}}(\mathbf{r}, \omega)=\mu_{0} \omega^{2} \mathbf{G}\left(\mathbf{r}, \mathbf{r}_{e}\right) \cdot \mathbf{p}_{c},
    \label{green_E}
\end{equation}
where $\mathbf{p}_{c} = 2\mathbf{p}$ is the classical dipole moment set to \myg{$\left|\mathbf{p}_{c}\right| = 1 \rm{D}$} \cite{ciraci2014}.
The field $\mathbf{E_{\rm{II}}}$ is found by solving a system of equations similar to Eqs. $\eqref{fq01}$, where $\mathbf{E}_0$ is substituted by the field associated with a $z$-oriented dipole radiating in free-space \cite{jurga2017}.

The radiative decay rate, $\gamma_{\mathrm{r}}$ can be found by evaluating the energy flow radiating out from the system:
\begin{equation}
    \gamma_{r}=\frac{1}{2} \frac{\gamma^{0}}{W^{0}} \int_{\partial \Omega} \operatorname{Re}\left\{\mathbf{E}_{\rm{II}} \times \mathbf{H}^{*}_{\rm{II}}\right\} \mathrm{d} A
\end{equation}
where $\gamma^{0} = \omega^{3}|\mathbf{p}|^{2} /\left(3 \hbar \pi \varepsilon_{0} c^{3}\right)$ is the radiative decay rate and $W^{0} = \omega^{4}|\mathbf{p}_c|^{2} /\left(12 \pi \varepsilon_{0} c^{3}\right)$ the radiated power in vacuum. Having found $\gamma_{\rm r}$, the nonradiative part is obtained as $\gamma_{\mathrm{nr}} = \gamma_{\mathrm{sp}} - \gamma_{\mathrm{r}}$, and the quantum yield $q$ can be found via:
\begin{equation}
    q = \frac{\gamma_{\mathrm{r}}}{\gamma_{\mathrm{sp}}}.
    \label{q_yield}
\end{equation}
Finally, the fluorescence enhancement of the emitter in the plasmonic environment of the dimer is given by:
\begin{equation}
    \eta_{em} = \eta_{ex}q.
    \label{fluor}
\end{equation}

In this article, we will consider the following cases with increasing complexity in the description of the electron dynamics in the Na spheres:

\begin{enumerate}[label=\Roman*:]
\item LRA, where the metal is described through the conventional Drude model, corresponding to $G\left[n\right]=0$ in Eq.~\eqref{fq01b};
\item TF-HT, in which we consider the TF pressure term in the hydrodynamic equation, i.e. $G\left[n\right]=T_{\rm TF}[n]$;
\item QHT, where we account for the vW KE and XC corrections,  $G\left[n\right]=T_{\rm TF}[n]+T_{\rm vW}[n,w]+ E_{\rm XC}[n]$, and consider a spatial dependent equilibrium charge density;
\item QHT-PGSL, in which the KE functional of the QHT is improved by adding the PGS and Laplacian terms, i. e., $\tau=\tau^{\mathrm{vW}}+\tau^{\mathrm{PGS}}+\tau^{\mathrm{TF}}\beta q_{r}^{2}$;
\item QHT-PGSLN, where we add the Laplacian-dependent logarithm correction to the QHT-PGSL, i. e., $\tau=\tau^{\mathrm{vW}}+\tau^{\mathrm{PGS}}+\tau^{\mathrm{TF}}\left[\beta q_{r}^{2}+2 \beta q_{0}^{2} \ln \left(1+q_{r} / q_{0}\right)\right]$.
\end{enumerate}

\section{Methods}\label{methods}
\noindent The system of Eqs.~\eqref{fq01}, where ${\bf E}_0$ is either a plane wave or a dipole filed, is solved using a finite element method through a customized implementation into a commercial software.
Moreover, to reduce the computational cost, we take advantage of the symmetry of the system, by expanding the fields in a series of cylindrical harmonics, each of which can propagate independently \cite{Ciraci:2013jt,khalid2019shell}.
Following this method, it is possible to reduce a three-dimensional calculation to few two-dimensional ones.
In fact, because the systems considered here are very small it is possible to limit the maximum azimuthal number $|m_{max}|\leq 1$.
Note that since the field associated with a vertically oriented dipole is axially symmetric, the emission properties can be exactly evaluated by only considering the azimuthal number $m = 0$ \cite{ciraciQEmmit}. More detail on the FEM implementation is given in the Appendix Section \ref{appfem}.

\section{Discussion of results}\label{discussion}
\noindent Let us start by comparing the enhancement of the incident field $\mathbf{E}_{0}$ inside the dimer gap when different models are considered.
\begin{figure}[t]
\centering
\resizebox{1\columnwidth}{!}{\includegraphics[width=1\textwidth,angle=0]{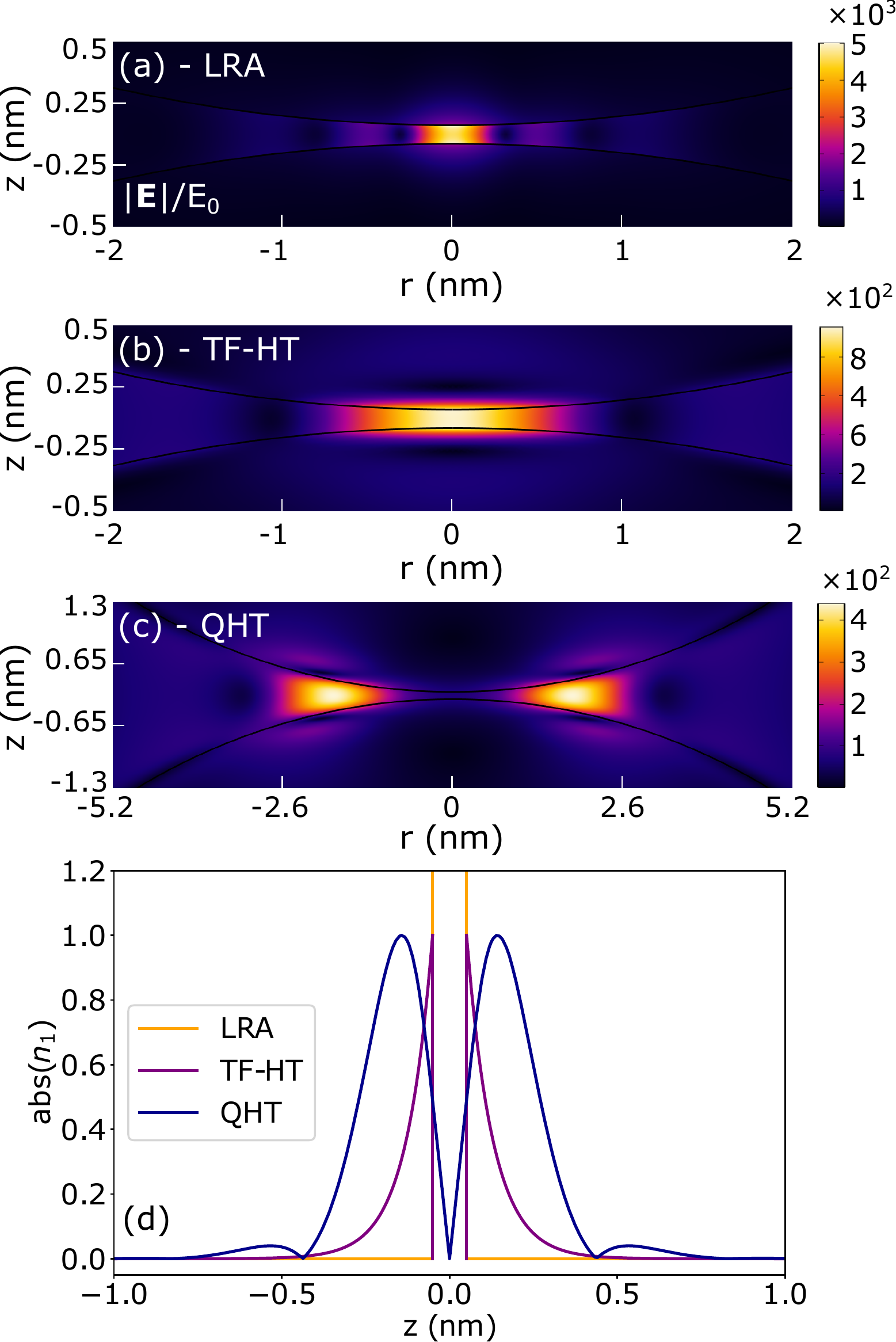}}
\caption{Norm of the total electric field (a-c) and the absolute value of the induced change density (d) as obtained from the LRA, TF and QHT at $\lambda = 400$~nm for \my{$g = 0.1$~nm} and $R = 10$~nm. The induced charge density is normalized to the maximum value in each scenario.}
\label{norm_n1}
\end{figure}
In Fig. \ref{norm_n1}, we show the distribution of the norm of the total field in the gap region for the LRA, TF-HT, and QHT approaches. Note that the spatial range is different for each model.
For the LRA (Fig. \ref{norm_n1} (a)), the field is extremely localized and greatly enhanced compared to other methods. This happens because the response of the system is purely local and the \my{induced charges can accumulate on an infinitesimally small layer at the metal surface confining the field in a much tighter region, as can be seen from Fig. \ref{norm_n1} (a)}. 
So, by shrinking the gap, we get increasing values for the field and the induced density, as can be seen in Fig. \ref{norm_n1} (d).
The situation is different for the TF-HT as depicted in Fig. \ref{norm_n1} (b). In this case, the \my{induced charge spreads into the bulk region} since the optical response is no longer completely local. Note that the field is localized into a volume of about 10 times wider than the previous case. For the TF-HT, we used $n_{0} = n^{+}$ and imposed hard-wall boundary conditions, i.e., $\mathbf{P} \cdot \hat{\mathbf{n}}=0$ \cite{ciraciQHT}, at the metal surface, which forces the induced charge density $n_1\left(\textrm{\textbf{r}}\right)$ to spread only inside the metal as shown in Fig. \ref{norm_n1} (d).

The most physical scenario is observed for the QHT in Fig. \ref{norm_n1} (c). \my{At optical frequencies, the electron spill-out from the metal surface \cite{zhu2016, esteban2012} results in the electron tunneling, which is the QHT case. This process is different to the one discussed in \cite{benz2015} where the tunneling happens through the emitter as the latter serves as a conductive linker .} Both the equilibrium density $n_{0}$ and the induced charge density $n_{1}$ are not constrained by the surface of the metal: we use Eq. \eqref{dens} for $n_0$ where the $s_{0}$ parameter controls the \my{spill-out, and hence the tunneling} of the density out of the metal and the hard-wall boundary condition on $n_1$ is lifted. As a result, the electric field goes into the metal and moves out of the gap center ($r = 0, z = 0$) by having maximum values in the regions to the left and right of the gap center.
Indeed, the behavior of the field resembles the movement of the liquid as it could be naturally foreseen considering the hydrodynamic approach to the problem (see Eq. \eqref{fq01b}). \my{The spill-out and tunneling affect the current density $\mathbf{J} = \partial \mathbf{P} / \partial t=-i\omega\mathbf{P}$, at the surface of the spheres for each scenario as it can be seen from Fig. S4 in SM.}

\begin{figure}[!hbt]
\centering
\includegraphics[width=0.44\textwidth,angle=0]{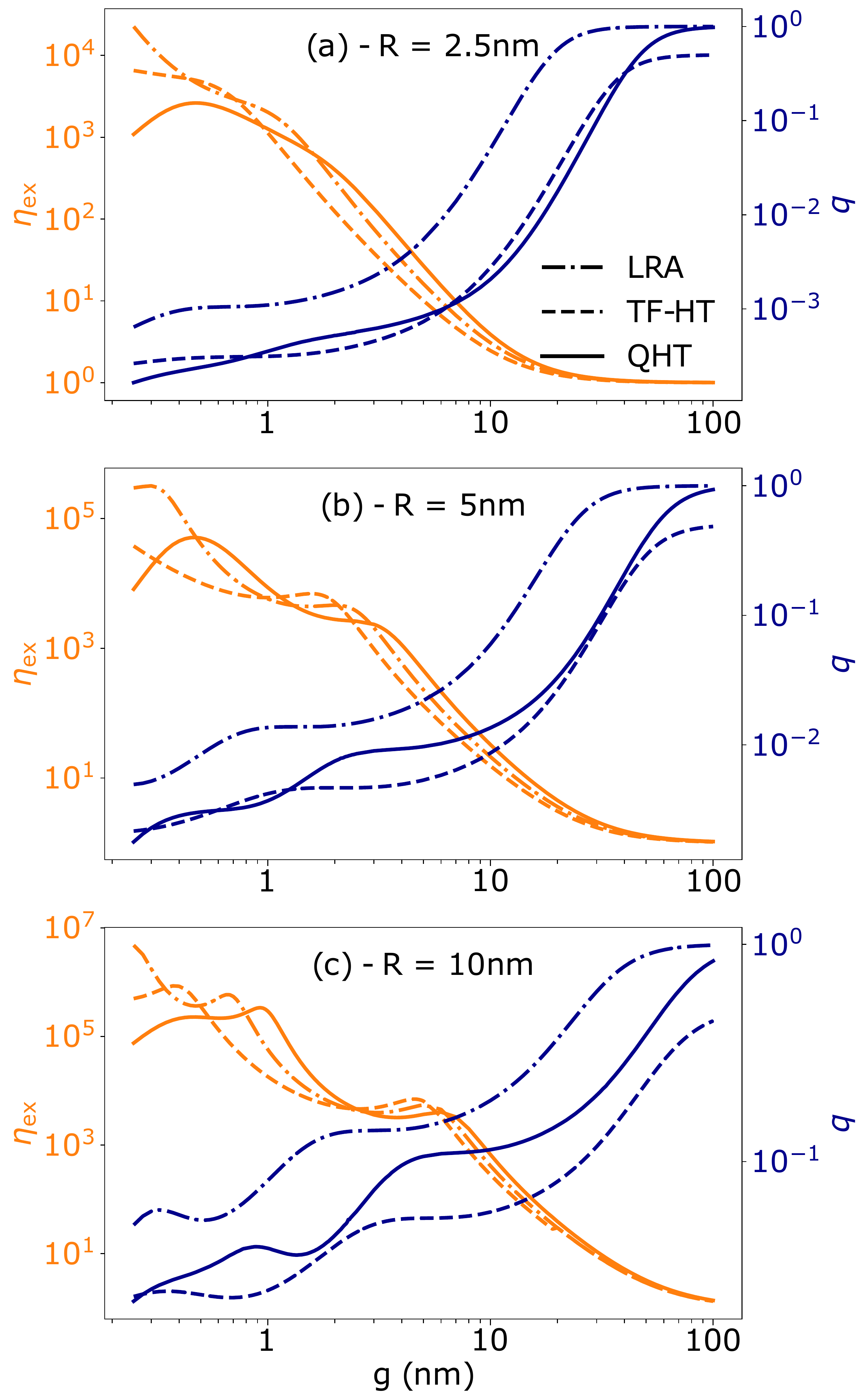}
\caption{\my{The dependence of the excitation enhancement $\eta_{ex}$ and quantum yield $q$ on the gap size (\myg{$0.25 \le g \le 100$~nm}) as calculated from the LRA, TF-HT and QHT for different radii at $\lambda = 400$~nm.}}
\label{ee_vs_q}
\end{figure}

Next, let us analyze how the terms in Eq.~\eqref{fluor}, i.e., the excitation enhancement, $\eta_{ex}$, and quantum yield, $q$, change by varying the size of the gap for the different models, as shown in Fig. \ref{ee_vs_q} for a fixed wavelength.
For large gaps, the electric field is localized only at the surface of the NPs (especially for the smaller systems) and, hence, $\eta_{ex}$ is relatively small and not much affected by the model used (the differences are due mostly to the shift of the plasmon resonance). 
\begin{figure*}[t]
\centering
{\includegraphics[width=1\textwidth]{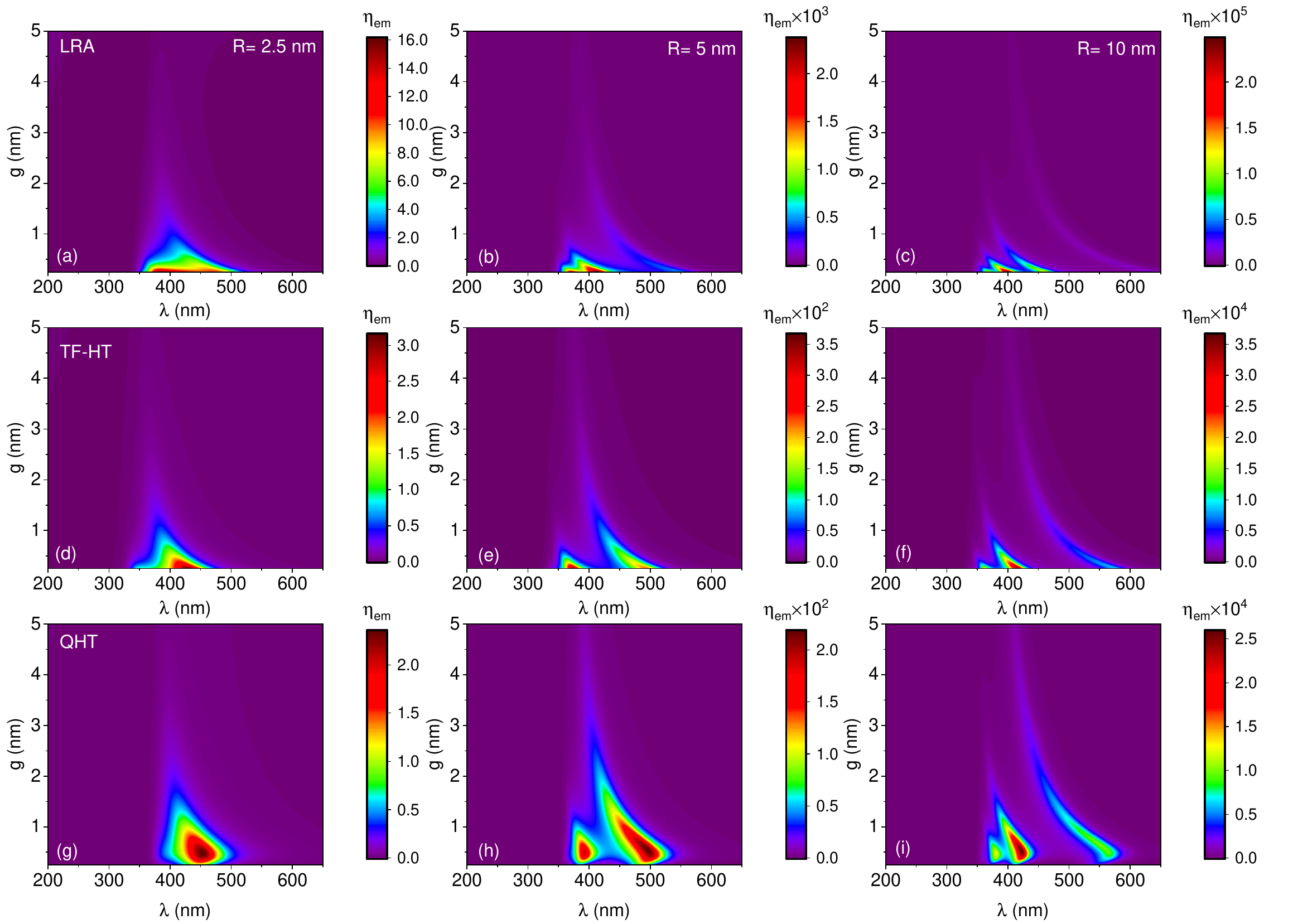}}
\caption{The fluorescence enhancement for different radii, $g$ and  $\lambda$ evaluated for the LRA, TF-HT and QHT.}
\label{fluor_map}
\end{figure*}

\begin{figure*}[t]
\centering
{\includegraphics[width=1\textwidth]{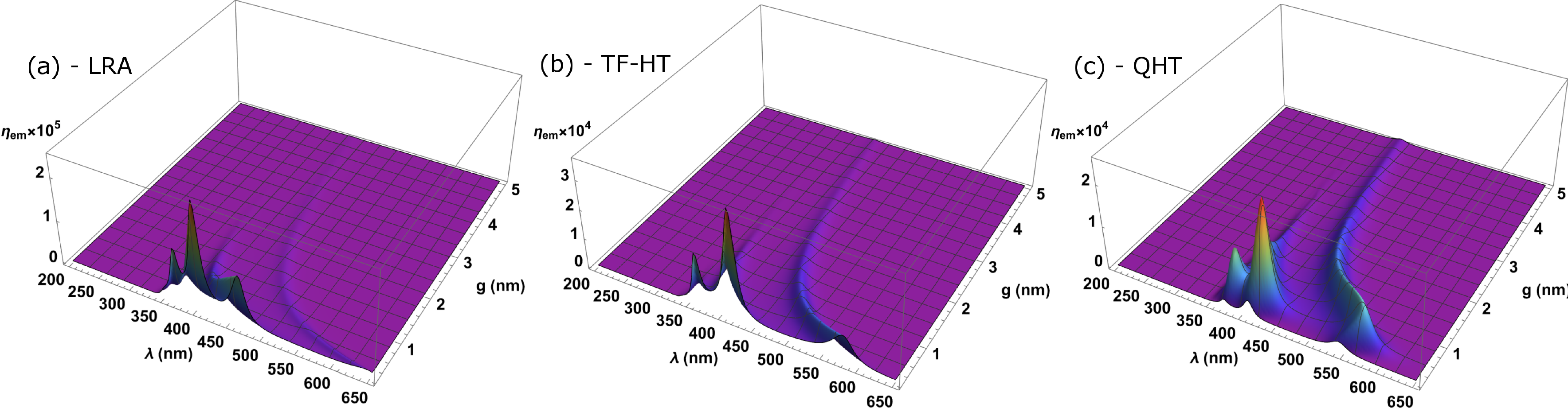}}
\caption{The dependence of the fluorescence enhancement for $R = 10$~nm gap size $g$ and the wavelength $\lambda$ as computed from LRA, TF-HT and QHT.}
\label{fluor_3D}
\end{figure*}
As the gap size shrinks, $\eta_{ex}$ reaches very high values for the LRA and TF-HT compared to the QHT as expected.
In addition, for QHT, we observe a drop at very small values of the gap since the field is mostly concentrated off the location of the emitter (see Fig. \ref{norm_n1} (c)).
An opposite behavior is observed for the quantum yield.
The smaller the gaps, the closer the emitter is to the NPs. 
At short distances, the evanescent field of the emitter can more easily coupled to plasmonic nonradiative modes channeling more energy toward Ohmic losses, and so reducing the quantum yield.
Note that the nonmonotonic behavior in $\eta_{ex}$ and $q$ is mainly due to the shift of the plasmon resonance as the gap varies.

Having seen singularly the behavior of $\eta_{ex}$ and $q$, we can now analyze the fluorescence enhancement shown in Fig. \ref{fluor_map} for the LRA, TF-HT and QHT, respectively.
For gaps $g>1$~nm, \my{our results are in line with previously published results for the LRA \cite{kongsuwan2018a}} and TF-HT \cite{Christensen:2014ha,Tserkezis:2016kp,jurga2017,ciraciQEmmit}.
Namely, the maximum fluorescence enhancement is found in the correspondence of the dipolar plasmon resonance (DPR).
At smaller gaps ($g\simeq0.5$~nm), high-order modes become more and more relevant in terms of fluorescence enhancement. 
This result however is not general and should be material-dependent. \my{We would like to mention that charge transfer modes can be observed in dimers, but they appear at energies below 1~eV \cite{barbry2015} and are not present in the spectral range shown in Figs. \ref{fluor_map} and \ref{fluor_3D}.}
As the gap closes, in fact, the DPR is shifted to much longer wavelengths for which Na ohmic losses increase, and both quantum yield and field enhancement reduce (see Fig. S1 and Fig. S2 in the Supplemental Material (SM).
This is clearly visible for the LRA, although the TF-HT and QHT show similar results but with more spread-out resonances due to the effect of nonlocality.
Note that for a fixed gap, increasing the NP radius has the same effect as reducing the gap  \cite{Christensen:2014ha}.
This is formally true for the LRA in the quasi-static limit, where spectra are exactly the same except for an overall factor (related to the size of the cavity) \cite{mayer2005, aubry2010}.

At $g \lesssim 0.5$~nm and for $R = 2.5$~nm, the maximum value of $\eta_{em}$ is at the dipolar plasmon resonance (DPR) for the TF-HT and QHT, but for the LRA, it is at the higher-order mode with a lower wavelength. If we remain in the same interval $g \lesssim 0.5$ but increase the radius, the maximum value gradually moves to the higher-order modes. These results mark the importance of considering the contribution of higher-order modes for smaller gaps, as it is also confirmed by TD-DFT calculations \cite{barbry2015}. At $g \gtrsim 0.5$, the main contribution to the fluorescence comes from the DPR for all cases.
In addition, the LRA predicts almost more than an order of magnitude higher values as compared to the TF-HT and QHT. Also, the LRA results in more peaks than the TF-HT and QHT for fixed radius. This could be attributed to the fact that the $\eta_{ex}$ is higher for the LRA, and hence higher-order modes can be distinguished from the DPR. As the radius increases, more higher-order modes appear in all cases \cite{deabajo1999}, and the DPR moves to the longer wavelengths. On the other hand, by widening the gap, the spheres become more separated, and the DPR moves to shorter wavelengths.

Let us now focus on the case with $R=10$~nm shown in Fig.~\ref{fluor_3D}.
The key difference of the QHT from the LRA and TF-HT is the quenching of fluorescence at gap sizes below $\approx 0.5$~nm.
This is somewhat an expected result from a physics standpoint since fluorescence cannot be arbitrarily enhanced.
The QHT then predicts the maximum fluorescence enhancement to occur at $g\approx 0.5$~nm before the electron tunneling starts to have a detrimental impact.  
Indeed, corrections introduced by the TF pressure are not enough to account for the quantum nature of the electrons at the metal surface, as it is confirmed by the monotonic increase of $\eta_{em}$ in Figs. \ref{fluor_3D}(a). In other words, just considering the nonlocality via electron pressure term still results in an overestimated fluorescence enhancement. \my{This has also been confirmed by considering spill-out and nonlocality via Feibelman $d$-parameters  \cite{goncalves2020}} and has resulted in the broadening of the Purcell enhancement spectra near the plasmon frequency in emitter Na sphere and emitter Na surface interactions.
The LRA, on the other hand, is likely to largely overestimate the attainable fluorescence enhancement at small gaps.
These result then indicates the necessity of considering both nonlocality and electron tunneling effects to correctly predict the fluorescence enhancement at sub-nanometer gaps.



\begin{figure}[!hbt]
\centering
\includegraphics[width=0.8\columnwidth]{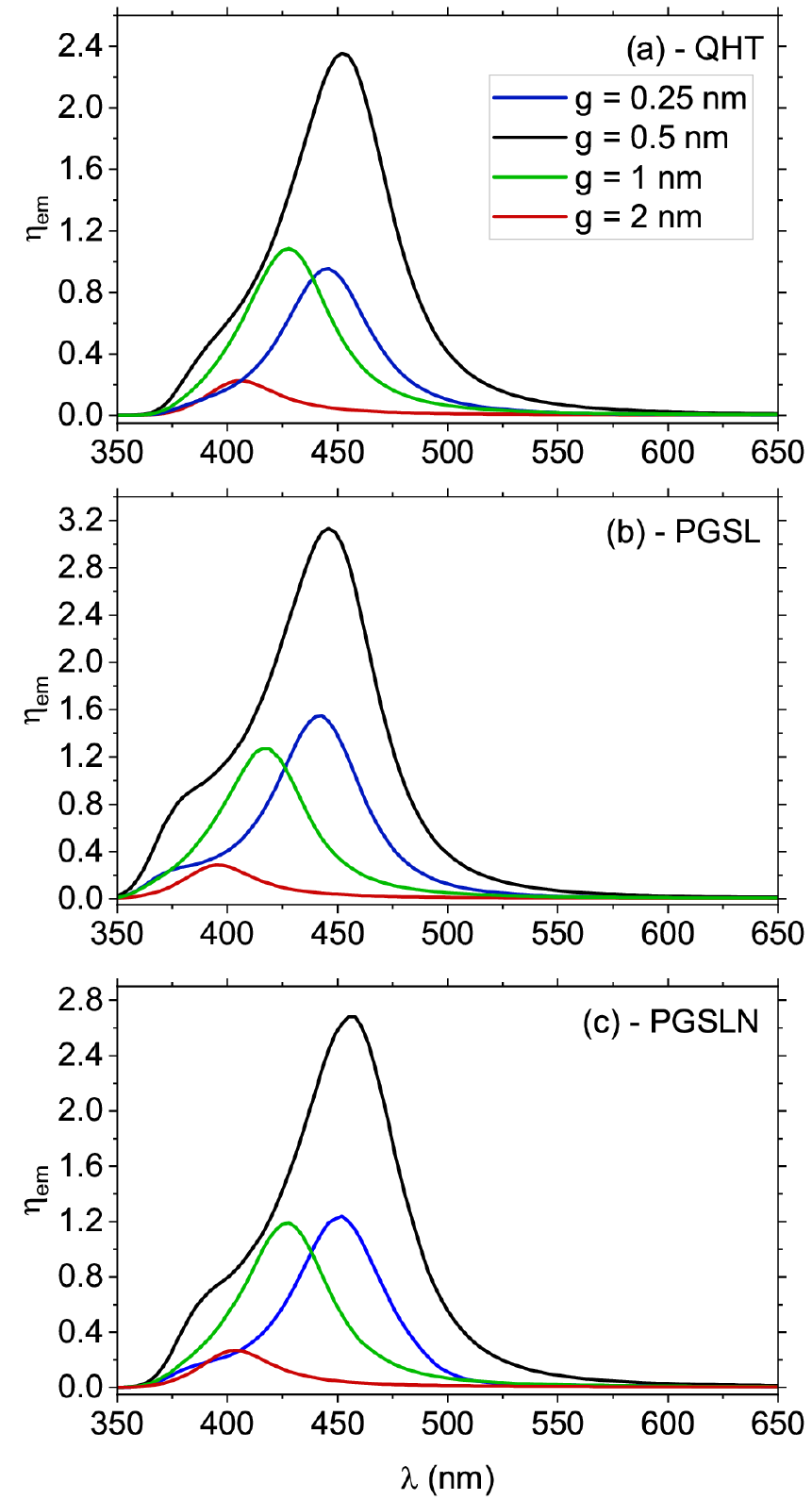}
\caption{The dependence of fluorescence enhancement on the wavelength for different gap values as calculated from the QHT, PGSL and PGSLN. The radius of spheres is $R = 2.5$~nm.}
\label{fluor_PGSLN}
\end{figure}

Finally, we compare the fluorescence enhancement as calculated from the QHT, QHT-PGSL, and QHT-PGSLN in Fig. \ref{fluor_PGSLN} for a sphere dimer with $R = 2.5$~nm for different values of the gap. \my{In our previous article \cite{baghramyan2021}, we showed that the PGSL and PGSLN do not give the nonphysical oscillations in the absorption spectrum of sphere dimers as compared to the QHT. Nevertheless, the contribution of those oscillations to the fluorescence spectra is very small compared to other modes. These oscillations should appear before $\lambda = 350$~nm in Figs. \ref{fluor_map} (g)-(i) and Fig. \ref{fluor_3D} (c) but clearly, we do not observe them. This is the reason why we choose the QHT over Laplacian models to compare with the LRA and TF-HT.}
It is interesting to observe that for all approaches the fluorescence is quenched as opposed to the LRA and TF-HT results.
This result again emphasizes the importance of considering the spill-out of the electron density at the metal surface and tunneling effects between closely spaced NPs, as it is incorporated in all quantum approaches (QHT, QHT-PGSL and QHT-PGSLN). 
As we can see, the biggest enhancement is achieved at $g =0.5$~nm for all scenarios and the highest value is obtained for the QHT-PGSL in Fig. \ref{fluor_PGSLN} (b).
The latter result is expected since the QHT-PGSL slightly overestimates the field inside the gap \cite{baghramyan2021}. 
On the other hand, the QHT-PGSLN does not suffer from such overestimation and, although more complex to implement, it should be considered as the most accurate.
Interestingly, these results do not differ much from the results obtained with the much simpler QHT model.

\section{Conclusions and Future Perspectives}\label{conculsions}

\noindent To conclude, we demonstrated that the consideration of nonlocality and tunneling in the framework of the QHT is crucial for the correct evaluation of fluorescence enhancement in plasmonic sphere dimers.
\my{While the LRA predicts increasing enhancement \cite{kongsuwan2018a,yang2020NatPhot}} as the NPs distance is decreased, the use of the QHT results in the quenching of the fluorescence of the emitter. In addition, only accounting for the nonlocality in the framework of the TF-HT still results in an increasing rate of fluorescence.
Results similar to the QHT are obtained if the second-order derivatives of the density are included in the definition of the KE functional associated to the electron dynamic in the metallic volumes in the form of PGSL and PGSLN functionals. The fluorescence quenching happens at approximately $g = 0.5$~nm for all scenarios. 
\my{These results should also be valid for similar systems, such as the NP on film \cite{kongsuwan2018a}.}

\myg{More generally, our method has allowed to introduce a fundamental mechanism for fluorescence quenching, which marks the limits for the local and hydrodynamic theories.
Even though our work is purely theoretical, experimental validation of such mechanism could be in principle achieved by using  analogous larger scale systems, such as a heavily-doped semiconductor systems, or by using single atom emitters at cryogenic temperatures, by exploiting the strong gradients of the electric field at the dimer gap in order to create an optical trap. Another possibility, is that of exploiting the weaker transverse dipole of traditional molecules, which would allow the molecule to lay flat in the gap, thus reaching distances below $\sim 0.5$ nm.}

Although we have considered only quantum effects in the dielectric response of the NPs, while assuming a point-dipole approximation for the emitter, we believe that equally important contribution could be introduced by taking into account the spatial extension of the quantum emitter beyond the point-dipole approximation \cite{Neuman:2018fl} and the quantum nature of light \cite{Feist.2020}.

\section*{Acknowledgments}
\noindent The authors thank  Antonio I. Fern\'andez-Dom\'inguez for insightful discussions and helpful comments.

\my{\appendix
\section{FEM implementation}\label{appfem}
We use FEM in COMSOL Multiphysics to solve the involved partial differential equations. Following the usual FEM procedure \eqref{fq01a} and \eqref{fq01b} are multiplied by the test functions $\textrm{\textbf{E}}_{\rm{I}}$ and $\textrm{\textbf{\~{P}}}$, and integration by parts is performed.
Also, since $\left(\frac{\delta G\left[n\right]}{\delta n}\right)_1$ can have fourth-order of $n_1$ derivatives for PGSL and PGSLN [see Expr. S5 in the SM], we introduce $\textrm{\textbf{F}} = \nabla n_1$ and $\textrm{\textbf{O}} = \nabla\left(\nabla^2 n_1\right) = \nabla^2\textrm{\textbf{F}}$ variables to lower the order to the first. In addition, since the geometry of dimer is axisymmetric, the \textrm{2.5D technique} is used \my{\cite{Ciraci:2013jt,ciraciQHT,ciraciFarfield2013}}. Following it,  the vector fields are expressed in cylindrical harmonics $v_{\rho, z, \phi}(\rho, z, \phi)=\sum_{m} v_{\rho, z, \phi}^{(m)}(\rho, z) e^{-i m \phi}$, where $\textrm{\textbf{v}} = \textrm{\textbf{E}}; \textrm{\textbf{P}};\textrm{\textbf{F}};\textrm{\textbf{O}}$, and $m \in \mathbb{Z}$. The test functions depend on $m$ via $e^{im\phi}$ form. Thus, we will have $2m_{\textrm{max}} + 1$ two-dimensional problems ($m_{\textrm{max}}$ is the maximum value for $m$), instead of the original three-dimensional problem. The final system of equations for all scenarios is the following:
}
\begin{widetext}
\begin{subequations}
    \begin{gather}
    \my{
        \int\left\{\left(\nabla\times\textrm{\textbf{E}}^{\left(m\right)}_{\rm{I}}\right)\cdot\left(\nabla\times\mathrm{\mathbf{\tilde{E}}}^{\left(m\right)}_{\rm{I}}\right)-\left(\frac{\omega^2}{c^2}\textrm{\textbf{E}}^{(m)}_{\rm{I}} + \mu_0\omega^2\textrm{\textbf{P}}^{(m)}\right)\cdot\mathrm{\mathbf{\tilde{E}}}^{\left(m\right)}_{\rm{I}}\right\}\rho d\rho dz = 0,} \label{aeq02a} \\
        \my{\int\left\{-\frac{e}{m_e}\left(\frac{\delta G\left[n\right]}{\delta n} \right)_1^{\left(m\right)}\left(\nabla \cdot \mathrm{\mathbf{\tilde{P}}}^{\left(m\right)} \right) + \frac{1}{n_0}\left[\left(\omega^2 + i\gamma\omega\right)\textrm{\textbf{P}}^{(m)} + \epsilon_0\omega_p^2\left(\textrm{\textbf{E}}^{(m)}_{\rm{I}} + \textrm{\textbf{E}}^{(m)}_\textrm{0}\right)\right]\cdot \mathrm{\mathbf{\tilde{P}}}^{\left(m\right)}\right\}\rho d\rho dz = 0,} \label{aeq02b} \\
        \my{\int\left\{\left(\nabla\cdot\textrm{\textbf{P}}^{\left(m\right)}\right)\left(\nabla\cdot\mathrm{\mathbf{\tilde{F}}}^{\left(m\right)}\right) + e\textrm{\textbf{F}}^{\left(m\right)}\cdot\mathrm{\mathbf{\tilde{F}}}^{\left(m\right)}\right\}\rho d\rho dz = 0,} \label{aeq02c}\\
        \my{\int\left\{\left(\nabla\cdot\textrm{\textbf{F}}^{\left(m\right)}\right)\left(\nabla\cdot\mathrm{\mathbf{\tilde{O}}}^{\left(m\right)}\right) + \textrm{\textbf{O}}^{\left(m\right)}\cdot\mathrm{\mathbf{\tilde{O}}}^{\left(m\right)}\right\}\rho d\rho dz = 0.} \label{aeq02d}
    \end{gather}
    \label{aeq02}
\end{subequations}
\end{widetext}

\my{For the LRA, we solved only the wave equation:}

\begin{widetext}
\begin{equation}
\my{\int\left\{\left(\nabla\times\textrm{\textbf{E}}^{\left(m\right)}_{\rm{I}}\right)\cdot\left(\nabla\times\mathrm{\mathbf{\tilde{E}}}^{\left(m\right)}_{\rm{I}}\right)-\frac{\omega^2}{c^2}\left(\textrm{\textbf{E}}^{(m)}_{\rm{I}} -  \left(\varepsilon_r-1\right) \textrm{\textbf{E}}^{(m)}_\textrm{0}\right)\cdot\mathrm{\mathbf{\tilde{E}}}^{\left(m\right)}_{\rm{I}}\right\}\rho d\rho dz = 0,}
\label{LRA_weq}
\end{equation}
\end{widetext}

\noindent \my{with Drude permittivity $\varepsilon = 1 - \frac{\omega_p^2}{\omega^2+i\omega\gamma}$. In the case of the TF-HT, Eqs. \eqref{aeq02a} and \eqref{aeq02b} need to be solved.}

\my{The similar procedure should be done for the evaluation of fields for decay processes, with the exception that now $\mathbf{E}_0$ is the $z$-oriented dipole field in free-space \cite{jurga2017}.}

\my{We used perfectly matched layers (PMLs) to emulate an infinite domain and avoid unwanted reflections, and a zero flux boundary condition is imposed on the electric field at the outer boundary of the PML. Also, the Dirichlet boundary conditions $\textrm{\textbf{P}} = 0, \textrm{\textbf{F}} = 0$, and $\textrm{\textbf{O}} = 0$ are set on the polarization domain (PD). For the TF-HT, an additional hard-wall boundary condition $\mathbf{P} \cdot \hat{\mathbf{n}}=0$ \cite{ciraciQHT} is set on the NP surface.}

\my{The meshing is done by using denser mesh on the surface of the dimer and on the artificial region around the dipole (see Figs. 3b, and 3c in the SM). We found that using a mapped mesh with rectangular elements at the location of the dipole results in a better convergence for the QHT, PGSL and PGSLN (see Fig. 3c in the SM). For other scenarios and other regions of the calculation domain we used only triangular meshing. For the PML, the mapped mesh with quadrilateral elements is used in all cases.}

\bibliography{bib_qht_flour}

\end{document}


\title{Supplementary Material for\\ ``Fluorescence quenching in plasmonic dimers due to electron tunneling''}

\author{Henrikh M. Baghramyan$^{1}$}
\author{Cristian Cirac\`{i}$^{1}$}
 \email{cristian.ciraci@iit.it}

\affiliation{$^{1}$Center for Biomolecular Nanotechnologies, Istituto Italiano di Tecnologia, Via Barsanti 14, 73010 Arnesano (LE), Italy}

\date{\today}

\maketitle
\def\mys{\color{black}}
\def\mye{\color{black}}
\newcommand{\my}[1]{\textcolor{black}{#1}}
\renewcommand{\thepage}{S\arabic{page}} 
\renewcommand{\thesection}{S\arabic{section}}  
\renewcommand{\thetable}{S\arabic{table}}  
\renewcommand{\thefigure}{S\arabic{figure}}
\renewcommand{\theequation}{S\arabic{equation}}

\my{\section{The excitation enhancement and the quantum yield}}

\begin{figure*}[!hbt]
\centering
{\includegraphics[width=1\textwidth,angle=0]{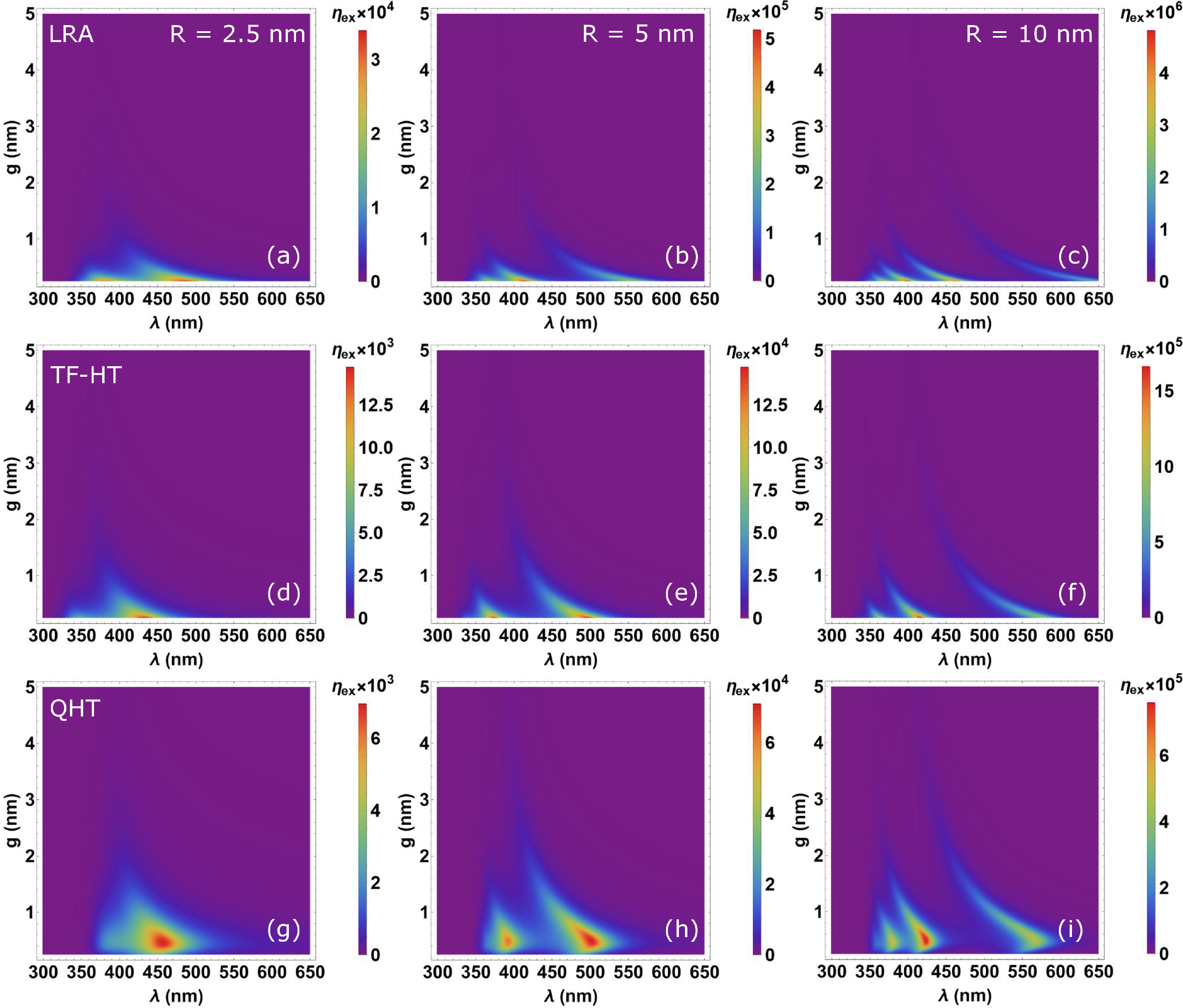}}
\caption{The dependence of the excitation enhancement on the gap size $g$ and the wavelength $\lambda$ as computed from LRA, TF-HT and QHT for $R = 2.5; 5; 10$~nm.}
\label{fluor_map}
\end{figure*}

\begin{figure*}[!hbt]
\centering
{\includegraphics[width=1\textwidth,angle=0]{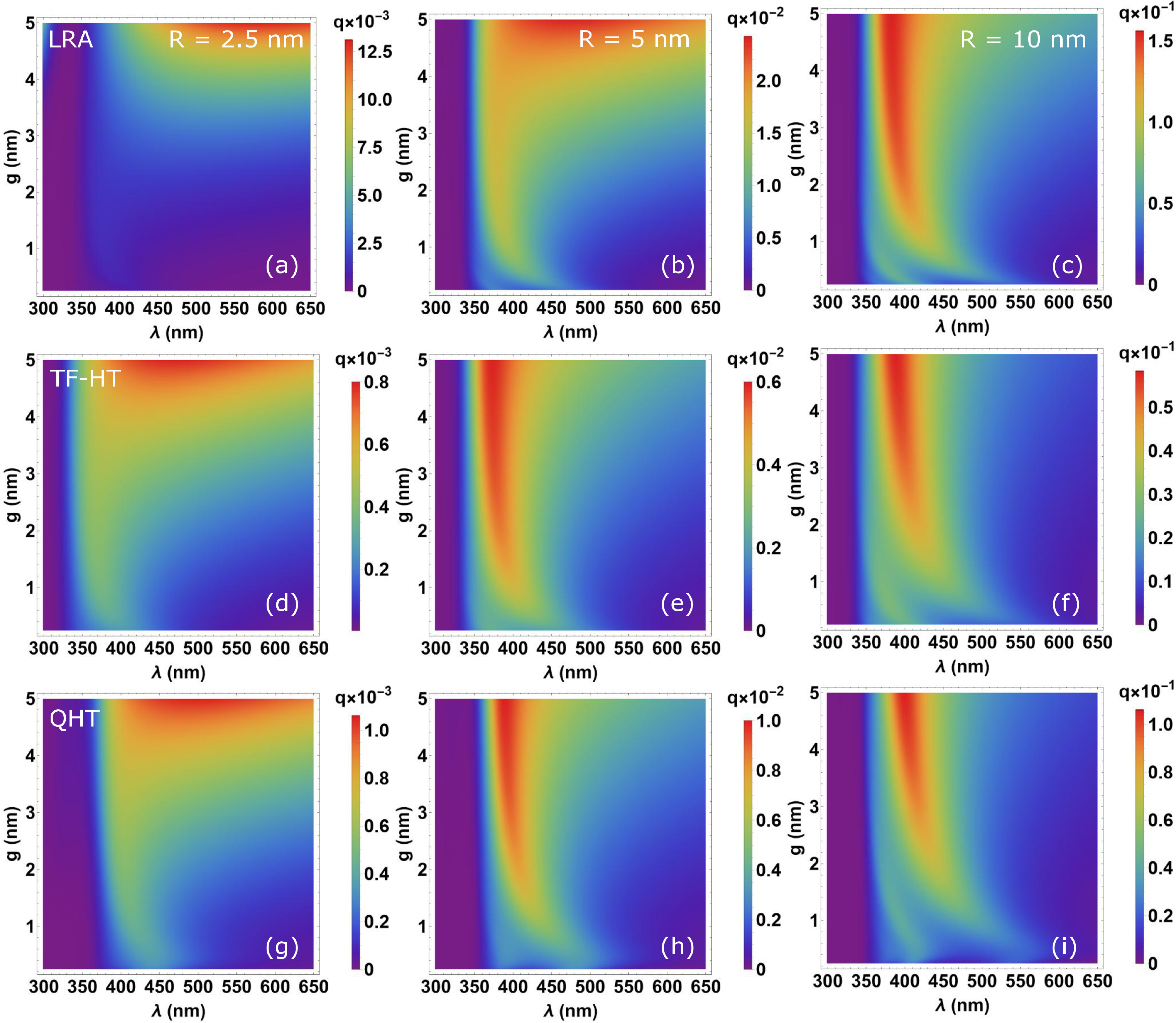}}
\caption{The dependence of the quantum yield on the gap size $g$ and the wavelength $\lambda$ as computed from LRA, TF-HT and QHT for $R = 2.5; 5; 10$~nm.}
\label{fluor_map}
\end{figure*}

\newpage

\my{\section{The computation domain and the meshing.}}

\begin{figure*}[!hbt]
\centering
{\includegraphics[width=0.8\textwidth,angle=0]{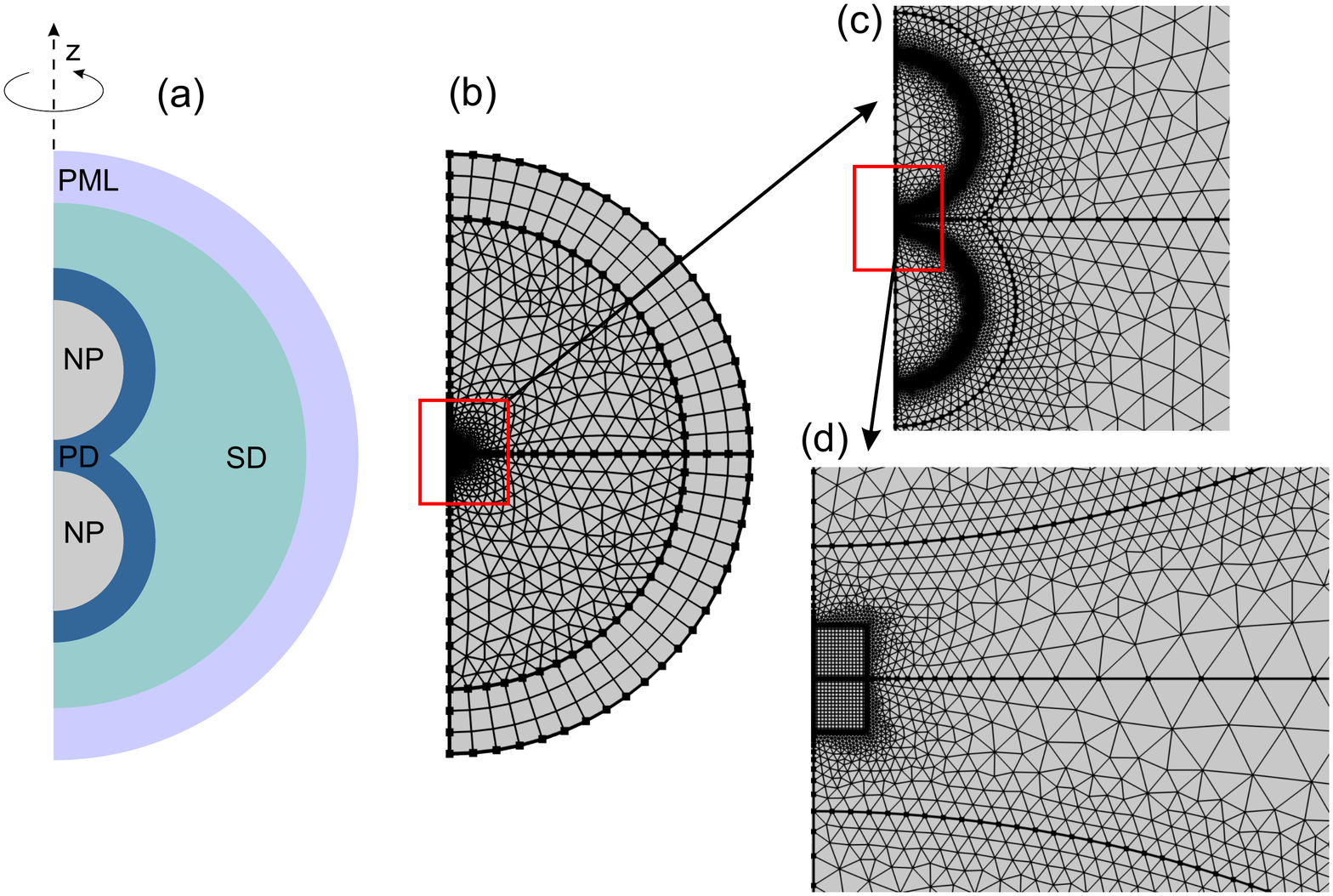}}
\caption{\my{The schematic images for the computation domain - (a), the meshing style used - (b) - (d).}}
\label{fluor_map}
\end{figure*}

\newpage

\my{\section{The current density.}}

\begin{figure*}[!hbt]
\centering
{\includegraphics[width=0.6\textwidth,angle=0]{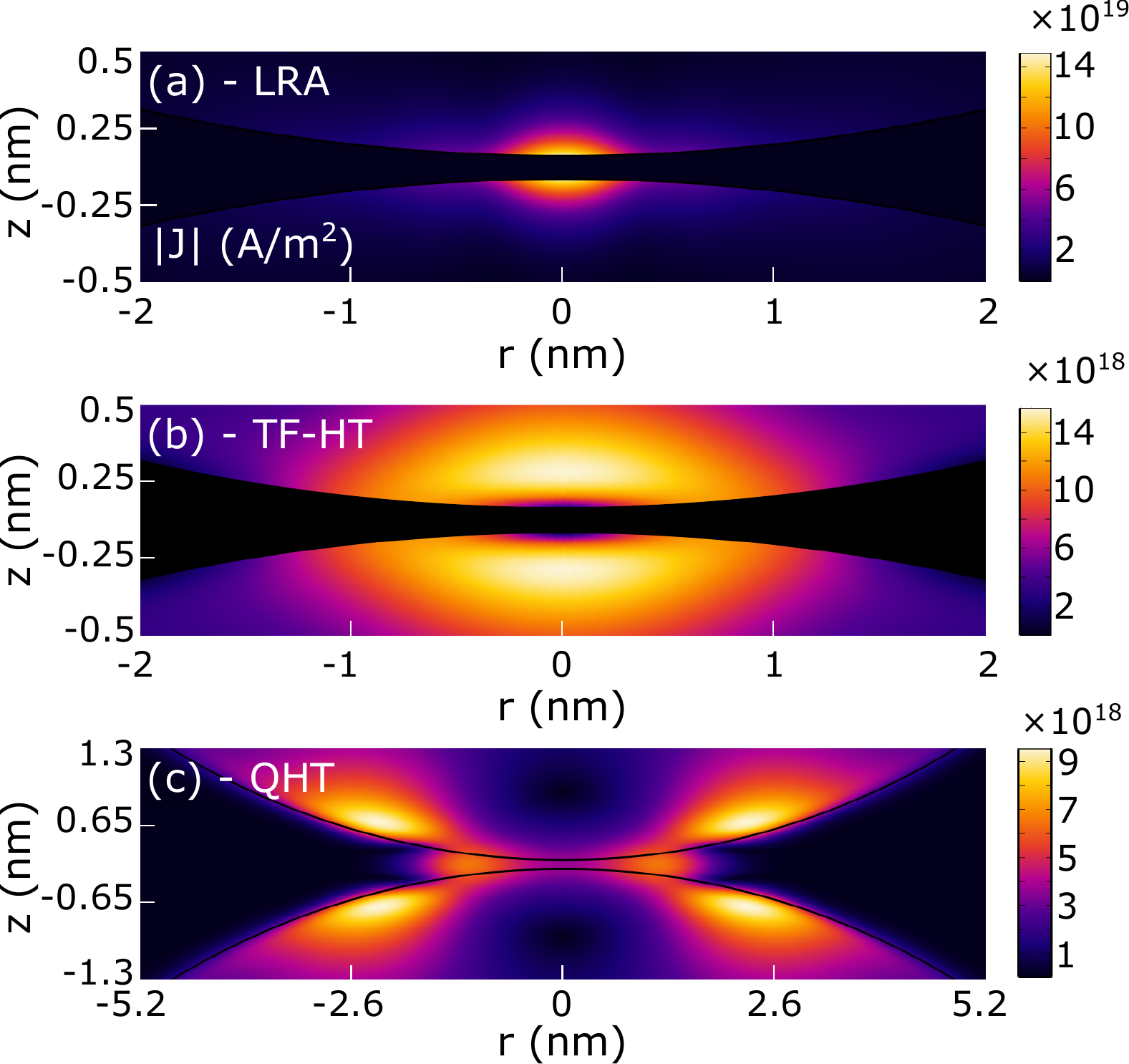}}
\caption{\my{Current density as obtained from LRA, TF and QHT at $\lambda = 400$~nm for $g = 0.1$~nm and $R = 10$~nm.}}
\label{Jdens}
\end{figure*}

\newpage

\my{\section{The potential}}

\my{The first-order terms in the potential} 

\my{\begin{equation}
     \left(\frac{\delta G\left[n\right]}{\delta n}\right)_1 =  \left(\frac{\delta E_{\mathrm{XC}}^{\mathrm{LDA}}}{\delta n}\right)_{1} + \left(\frac{\delta T_s}{\delta n}\right)_1
\end{equation}}

\my{are presented below.}

\my{For the exchange-correlation potential, it is given by:}

\my{\begin{equation}
\left(\frac{\delta E_{\mathrm{XC}}^{\mathrm{LDA}}}{\delta n}\right)_{1}=\left(E_{h}\right)\left(-a_{0} \frac{4}{9} c_{X} n_{0}^{-2 / 3} n_{1}+a_{0}^{3} \mu_{C}^{\prime}\left[n_{0}\right] n_{1}\right),
\label{EXC_pot}
\end{equation}}

\noindent \my{where} 

\my{\begin{equation}
\mu_{C}^{\prime}[n]= \begin{cases}-\frac{4 \pi}{9}\left[a+\frac{1}{3}\left(1+2 \ln \left(r_{s}\right)\right) c r_{s}+\frac{2}{3} d r_{s}\right] r_{s}^{3}, & r_{s}<1, \\ \frac{\alpha \pi}{27} \frac{5 \beta_{1} \sqrt{r_{s}}+\left(7 \beta_{1}^{2}+8 \beta_{2}\right) r_{s}+21 \beta_{1} \beta_{2} r_{s}^{3 / 2}+16 \beta_{2}^{2} r_{s}^{2}}{\left(1+\beta_{1} \sqrt{r_{s}}+\beta_{2} r_{s}\right)^{3}}, & r_{s} \geqslant 1 ,\end{cases}
\end{equation}}

\noindent\my{with $c_{X}=\frac{3}{4}\left(\frac{3}{\pi}\right)^{1 / 3}$ and the values of the coefficients are $a=0.0311$, $b=-0.048$, $c=0.002$, $d=-0.0116$, $\alpha=-0.1423, \beta_{1}=1.0529$, and $\beta_{2}=0.3334$.}

\my{The expression for $\left(\frac{\delta T_s}{\delta n}\right)_1$ is:}

\my{\begin{equation}
        \left(\frac{\delta T_s}{\delta n}\right)_1 = 
          \left(\frac{\delta T_s^{\textrm{I}}}{\delta n}\right)_1
        + \left(\frac{\delta T_s^{\textrm{II}}}{\delta n}\right)_1 \\ 
        + \left(\frac{\delta T_s^{\textrm{III}}}{\delta n}\right)_1,
        \label{fq07}
\end{equation}}
\my{where}
\my{\begin{widetext}
    \begin{subequations}
        \begin{align}
            \left(\frac{\delta T_s^{\textrm{I}}}{\delta n}\right)_1 = & \, \tau_{nn}^{(0)} n_1 \label{eqf:1}\\
            \left(\frac{\delta T_s^{\textrm{II}}}{\delta n}\right)_1 = & - 2\tau_{nnw}^{(0)}\left|\nabla n_0\right|^2n_1 - 2\tau_{nw}^{(0)}\left[n_1 \nabla^2 n_0 + \nabla n_0 \cdot \nabla n_1\right] - 2\tau_w^{(0)} \nabla^2 n_1  \label{eqf:2}\\  
            & -2\tau_{ww}^{(0)} \left[2\left(\nabla n_0 \cdot \nabla n_1\right)\nabla^2 n_0 + \nabla \left(\left|\nabla n_0\right|^2 \right) \cdot \nabla n_1 + 2\nabla n_0 \cdot \nabla \left(\nabla n_0 \cdot \nabla n_1\right)\right]  \label{eqf:3} \\ 
                    & - 4\tau_{www}^{(0)} \left(\nabla n_0 \cdot \nabla n_1 \right)\left[\nabla n_0 \cdot \nabla \left(\left|\nabla n_0\right|^2 \right) \right] \label{eqf:4} \\
                    & -2\tau_{nww}^{(0)} \left[2\left(\nabla n_0 \cdot \nabla n_1 \right)\left|\nabla n_0\right|^2 + \left\{\nabla n_0 \cdot \nabla \left(\left|\nabla n_0\right|^2 \right) \right\}n_1\right], \label{eqf:5} \\
           \left(\frac{\delta T_s^{\textrm{III}}}{\delta n}\right)_1 = & \,
           \tau_{nnnq}^{(0)}\left|\nabla n_0\right|^2  n_1 + \tau_{nnq}^{(0)} \left[2\nabla n_0 \cdot \nabla n_1 + n_1 \nabla^2 n_0 \right] + 2\tau_{nq}^{(0)} \nabla^2 n_1  \label{eqf:6} \\
                    & + \tau_{nnqq}^{(0)} \left[\left|\nabla n_0\right|^2 \nabla^2 n_1 + 2 n_1\left\{\nabla n_0 \cdot \nabla\left(\nabla^2 n_0\right)\right\}\right]  \label{eqf:7} \\
                    & + \tau_{nqq}^{(0)} \left[\nabla^2 n_0 \nabla^2 n_1 + 2 \nabla n_1 \cdot \nabla\left(\nabla^2 n_0\right) + 2\nabla n_0 \cdot \nabla \left(\nabla^2 n_1\right) + \nabla^2\left(\nabla^2n_0\right)n_1\right] \label{eqf:8} \\
                    & +  \tau_{qq}^{(0)} \nabla^2\left(\nabla^2n_1\right) \label{eqf:9}  \\
              & + \tau_{q q q}^{(0)}\left[2\left(\nabla\left(\nabla^{2} n_{0}\right) \cdot \nabla\left(\nabla^{2} n_{1}\right)\right)+\nabla^{2}\left(\nabla^{2} n_{0}\right) \nabla^{2} n_{1}\right]  \label{eqf:10} \\
                    & +\tau_{n q q q}^{(0)}\left[2\left\{\nabla n_{0} \cdot \nabla\left(\nabla^{2} n_{0}\right)\right\} \nabla^{2} n_{1}+\left|\nabla\left(\nabla^{2} n_{0}\right)\right|^{2} n_{1}\right]  \label{eqf:11} \\
                    & +\tau_{q q q q}^{(0)}\left|\nabla\left(\nabla^{2} n_{0}\right)\right|^{2} \nabla^{2} n_{1}  \label{eqf:12}.
            \end{align}
            \label{fq08}
    \end{subequations}
\end{widetext}}
\my{The subscripts  $i = n, w, q$ denote the partial derivatives, and the superscript $(0)$ means that the function is evaluated at $n = n_0$.}

\my{The potentials for each scenario considered in the article are as follows.}

\my{\begin{itemize}
  \item For LRA there is no functional and hence $\left(\frac{\delta G\left[n\right]}{\delta n}\right)_1=0$.
  \item For TF-HT $\left(\frac{\delta G\left[n\right]}{\delta n}\right)_1$ is given by the Eq. \eqref{eqf:1}.
  \item For QHT, Eqs. \eqref{eqf:1} and \eqref{eqf:2} are considered.
  \item For QHT-PGSL we use Eqs. \eqref{eqf:1}-\eqref{eqf:9}.
  \item And finally, for QHT-PGSLN all the equations in \eqref{fq08} have to be accounted for.
\end{itemize}}

\clearpage